\documentclass[runningheads,a4paper]{llncs}

\usepackage{amssymb,amsmath}
\usepackage{graphicx,subfig}

\usepackage{url}
\urldef{\mailsa}\path|szzoli@cs.elte.hu|
\newcommand{\keywords}[1]{\par\addvspace\baselineskip
\noindent\keywordname\enspace\ignorespaces#1}

\usepackage{color}            
\definecolor{mygreen}{rgb}{.1,.5,.1}
\definecolor{myblue}{rgb}{.1,.1,.8}
\usepackage[hyperfootnotes=false,
colorlinks=true,                 
linkcolor=myblue,                
urlcolor=myblue,                 
citecolor=mygreen,               
dvips]{hyperref}

\newcommand{\bof}{\mathbf} 

\newcommand{\R}{\mathbb{R}}
\begin{document}

\mainmatter 

\title{Towards Nonstationary, Nonparametric Independent Process Analysis with Unknown Source Component Dimensions}
\titlerunning{Nonparametric Independent Process Analysis}

\author{Zolt\'{a}n Szab\'{o}}
\institute{E\"{o}tv\"{o}s Lor{\'a}nd University, Department of Software Technology and Methodology\\
P\'{a}zm\'{a}ny P{\'e}ter s{\'e}t{\'a}ny 1/C, Budapest, H-1117, Hungary\\
\mailsa\\
\url{http://nipg.inf.elte.hu/szzoli}}

\maketitle

\begin{abstract}
The goal of this paper is to extend independent subspace analysis (ISA) to the case of (i) nonparametric, not strictly stationary source dynamics and (ii) unknown source component dimensions. We make use of functional autoregressive (fAR) processes to model the temporal evolution of the hidden sources. An extension of the ISA separation principle--which states that the ISA problem can be solved by traditional independent component analysis (ICA) and clustering of the ICA elements--is derived for the solution of the defined fAR independent process analysis task (fAR-IPA): applying fAR identification we reduce the problem to ISA. A local averaging approach, the Nadaraya-Watson kernel regression technique is adapted to obtain strongly consistent fAR estimation. We extend the Amari-index to different dimensional components and illustrate the efficiency of the fAR-IPA approach by numerical examples.
\end{abstract}

\section{Introduction}\label{sec:introduction}
Independent Component Analysis (ICA) \cite{jutten91blind,comon94independent} has received considerable attention
in signal processing. One may consider ICA as a cocktail party problem: we have $D$ speakers
(sources) and $D$ microphones (sensors), which measure the mixed signals emitted by the sources.
The task is to recover the original sources from the mixed observations only. For a recent review about ICA, see \cite{cichocki02adaptive}. In ICA the hidden independent sources are one-dimensional. The model is
more realistic if we assume that not all, but only some groups of the hidden sources are
independent (`speakers are talking in groups'). This is the Independent Subspace Analysis (ISA)
generalization of the ICA problem \cite{cardoso98multidimensional}. The ISA model has already had some exciting applications including  (i) the analysis of EEG, fMRI, ECG signals and gene data, (ii) pattern and face direction
recognition. For a recent review of ISA techniques, see \cite{szabo09separation}.

One can relax the traditional independent identically distributed (i.i.d.) assumption of ISA and model the temporal evolution of the sources, for example, by autoregressive (AR) processes \cite{poczos05independent}, however to the best of our knowledge, the general case of sources with unknown, nonparametric dynamics has been hardly touched in the literature \cite{theis05blind,anemuller06independent}. \cite{anemuller06independent} focused on the separation of stationary and ergodic source components of known and equal dimensions, in case of constrained mixing matrices. \cite{theis05blind} was dealing with wide sense stationary sources that (i) are supposed to be block-decorrelated for all time-shifts, and (ii) have equal and known dimensional source components. 

One of the most exciting and fundamental hypotheses of the ICA research is due to Jean-Fran\c{c}ois Cardoso, who conjectured that the solution of the ISA problem can be separated \cite{cardoso98multidimensional} to (i) applying traditional ICA and then (ii) clustering of the ICA elements into statistically dependent groups. While the extent of this conjecture, the \emph{ISA separation principle} is still on open issue, it has been rigorously proven for some distribution types \cite{szabo07undercomplete_TCC}. The goal of the present paper is to address the problem of ISA with nonparametric dynamics.
Beyond the extension to not necessarily stationary dynamics, the temporal evolution of the sources can be coupled (it is sufficient that their driving noises are independent, no block-decorrelatedness for the sources are required) and we treat the case of unknown source component dimensions. We model the dynamics of the sources by functional AR (fAR) processes and derive separation principle based solution for the resulting problem: the task is transformed to fAR estimation and ISA. To obtain strongly consistent fAR estimation the Nadaraya-Watson kernel regression technique is invoked.

The paper is structured as follows: Section~\ref{sec:fAR-IPA model} formulates the problem domain.
Section~\ref{sec:method} shows how to transform the problem to functional AR estimation task and ISA, and presents the kernel regression based approach. Section~\ref{sec:illustrations} contains the numerical illustrations. Conclusions are drawn in
Section~\ref{sec:conclusions}.

\section{The Functional Autoregressive Independent Process Analysis Model}\label{sec:fAR-IPA model}
We define the functional autoregressive independent process analysis (fAR-IPA) model. Let us assume that the observation ($\bof{x}$) is linear mixture ($\bof{A}$) of the hidden source ($\bof{s}$), which evolves according to an unknown fAR dynamics ($\bof{f}$) with independent driving noises ($\bof{e}$). Formally,
\begin{align}
 \bof{s}_{t} &= \bof{f}(\bof{s}_{t-1},\ldots,\bof{s}_{t-p})+\bof{e}_t,\label{eq:fAR-IPA:1}\\
 \bof{x}_t &= \bof{As}_t,\label{eq:fAR-IPA:2}
\end{align}
where the unknown mixing matrix $\bof{A}\in\R^{D\times D}$ is invertible, $p$ is the order of the process and the $\bof{e}^m\in\R^{d_m}$ components of $\bof{e}=\left[\bof{e}^1;\ldots;\bof{e}^M\right]\in\R^{D}$ $(D=\sum_{m=1}^Md_m)$ are (i) non-Gaussian, (ii) i.i.d.\ in time $t$ and (iii) independent, $I(\bof{e}^1,\ldots,\bof{e}^M)=0$, where $I$ denotes mutual information. The goal of the fAR-IPA problem is
to estimate (i) the inverse of mixing matrix $\bof{A}$, $\bof{W} =\bof{A}^{-1}$ and (ii) the original source $\bof{s}_t$ by using observations $\bof{x}_t$ only.

\section{Method}\label{sec:method}
The estimation of the fAR-IPA problem \eqref{eq:fAR-IPA:1}-\eqref{eq:fAR-IPA:2} can be accomplished as follows. The observation process $\bof{x}$ is invertible linear transformation of the hidden fAR source process $\bof{s}$ and thus it is also fAR process with innovation $\bof{Ae}_t$
\begin{align}
 \bof{x}_t &= \bof{As}_t = \bof{Af}(\bof{s}_{t-1},\ldots,\bof{s}_{t-p})+\bof{Ae}_t=\\
	 &= \bof{Af}(\bof{A}^{-1}\bof{x}_{t-1},\ldots,\bof{A}^{-1}\bof{x}_{t-p})+\bof{Ae}_t=\\
	 & = \bof{g}(\bof{x}_{t-1},\ldots,\bof{x}_{t-p})+\bof{n}_t,\label{eq:fAR-IPA:2b}
\end{align}
where function $\bof{g}(\bof{u}_{1},\ldots,\bof{u}_{p})=\bof{Af}(\bof{A}^{-1}\bof{u}_{1},\ldots,\bof{A}^{-1}\bof{u}_{p})$ describes the temporal evolution of $\bof{x}$ and $\bof{n}_t=\bof{Ae}_t$ stands for the driving noise of the observation. Making use of this form, the fAR-IPA estimation can be carried out by fAR fit to observation $\bof{x}$ followed by ISA on $\hat{\bof{n}}_t$, the estimated innovation of $\bof{x}$.

Let us notice that Eq.~\eqref{eq:fAR-IPA:2b} can be considered as a nonparametric regression problem, we have $\bof{u}_t=[\bof{x}_{t-1},\ldots,\bof{x}_{t-p}]$, $\bof{v}_t=\bof{x}_t$ $(t=1,\ldots,T)$ samples from the unknown relation
\begin{equation}
 \bof{v}_t = \bof{g}(\bof{u}_t)+\bof{n}_t,  
\end{equation}
where $\bof{u}$, $\bof{v}$ and $\bof{n}$ is the explanatory-, response variable and noise, respectively, and $\bof{g}$ is the unknown conditional mean or regression function. Nonparametric techniques can be applied to estimate the unknown mean function $\bof{g}( \bof{U})=E(\bof{V}|\bof{U})$, e.g., by carrying out kernel density estimation for random variables ($\bof{u}$,$\bof{v}$) and $\bof{u}$, where $E$ stands for expectation. The resulting Nadaraya-Watson estimator (i) takes the simple form
\begin{equation}
\hat{\bof{g}}_0(\bof{u})=\frac{\sum_{t=1}^T\bof{v}_tK\left(\frac{\bof{u}-\bof{u}_t}{h}\right)}{\sum_{t=1}^TK\left(\frac{\bof{u}-\bof{u}_t}{h}\right)},
\end{equation}
where $K$ and $h>0$ denotes the applied kernel (a non-negative real-valued function that integrates to one) and bandwith, respectively, and (ii) can be used to provide a strongly consistent estimation of the regression function $\bof{g}$ for stationary $\bof{x}$ processes \cite{bosq98nonparametric}. It has been shown recently \cite{hilgert09strong}, that for first order and only asymptotically stationary fAR processes, under mild regularity conditions, one can get strongly constistent estimation for innovation $\bof{n}$ by applying the recursive version of the Nadaraya-Watson estimator
\begin{equation}
\hat{\bof{g}}(\bof{u}) = \frac{\sum_{t=1}^Tt^{\beta D} \bof{v}_t K(t^{\beta}(\bof{u}-\bof{u}_t))}{\sum_{t=1}^Tt^{\beta D} K(t^{\beta}(\bof{u}-\bof{u}_t))},\label{eq:recNW}
\end{equation}
where the bandwith is parameterized by $\beta\in(0,1/D)$.

\section{Illustrations}\label{sec:illustrations}
Now we illustrate the efficiency of the algorithm presented in
Section~\ref{sec:method}. Test databases are described
in Section~\ref{sec:databases}. To evaluate the solutions we use the
performance measure given in Section~\ref{sec:amari-index}. The
numerical results are summarized in Section~\ref{sec:simulations}.

\subsection{Databases}\label{sec:databases}
\begin{figure}%
\centering%
\subfloat[][]{\includegraphics[height=2cm]{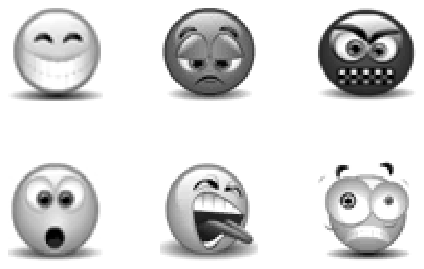}}\hfill%
\subfloat[][]{\includegraphics[height=2cm]{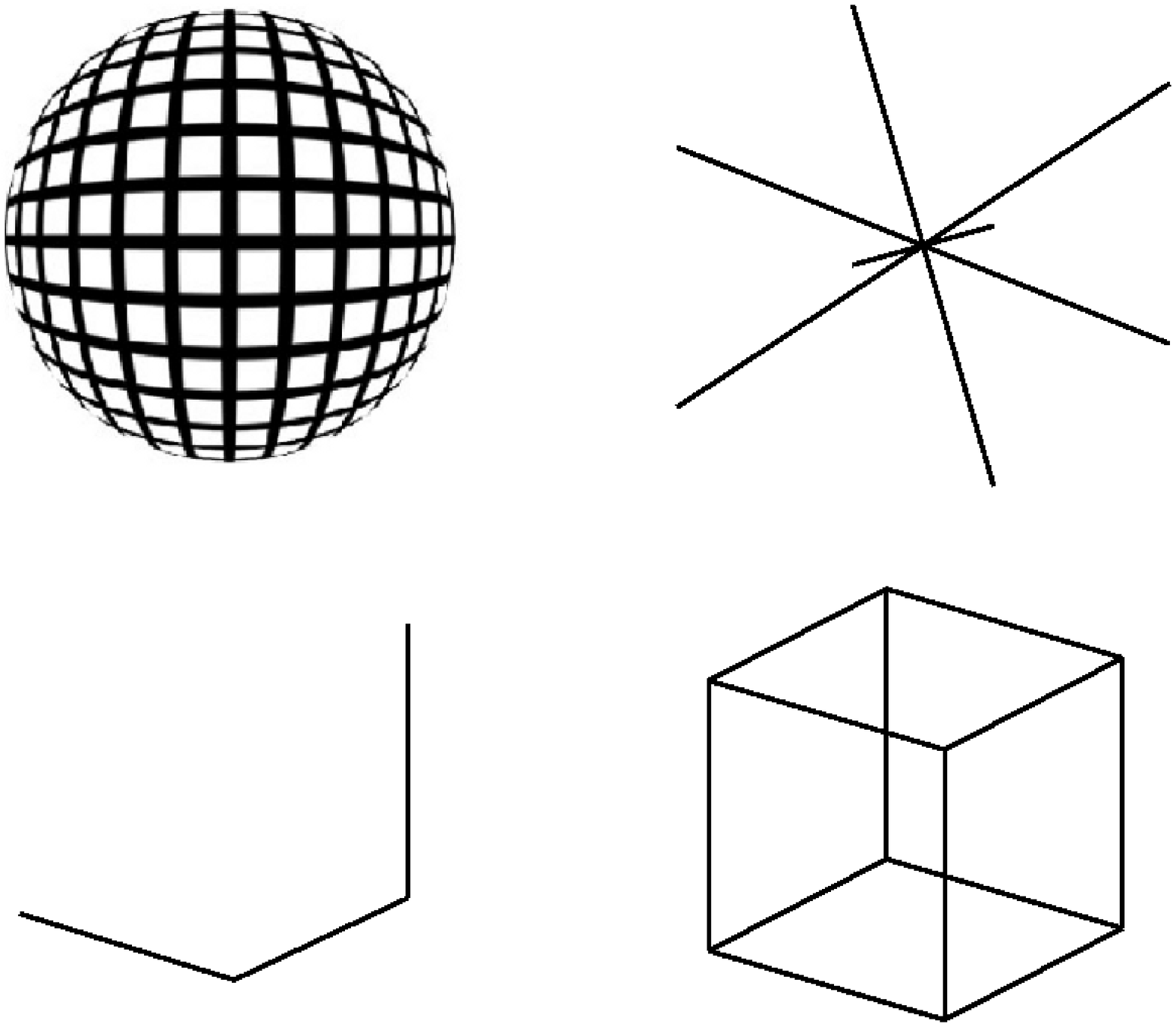}}\hfill
\subfloat[][]{\includegraphics[height=2cm]{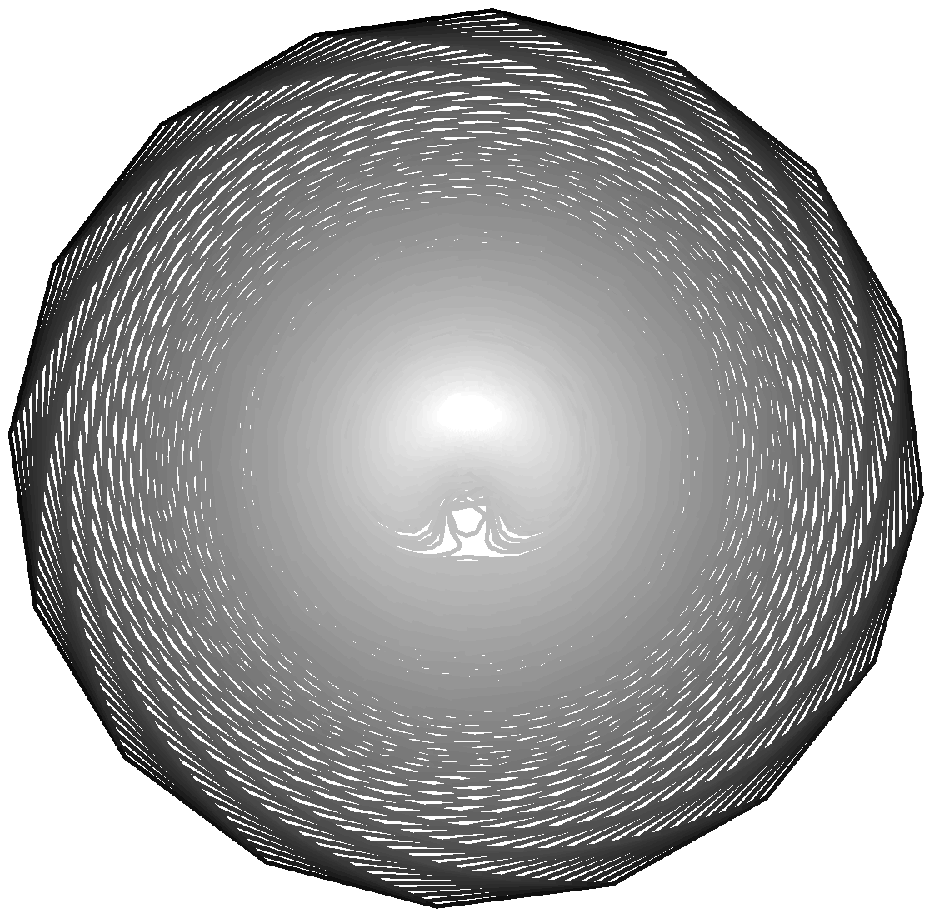}\hspace*{0.25cm}
\vspace*{-1cm}\includegraphics[height=2cm]{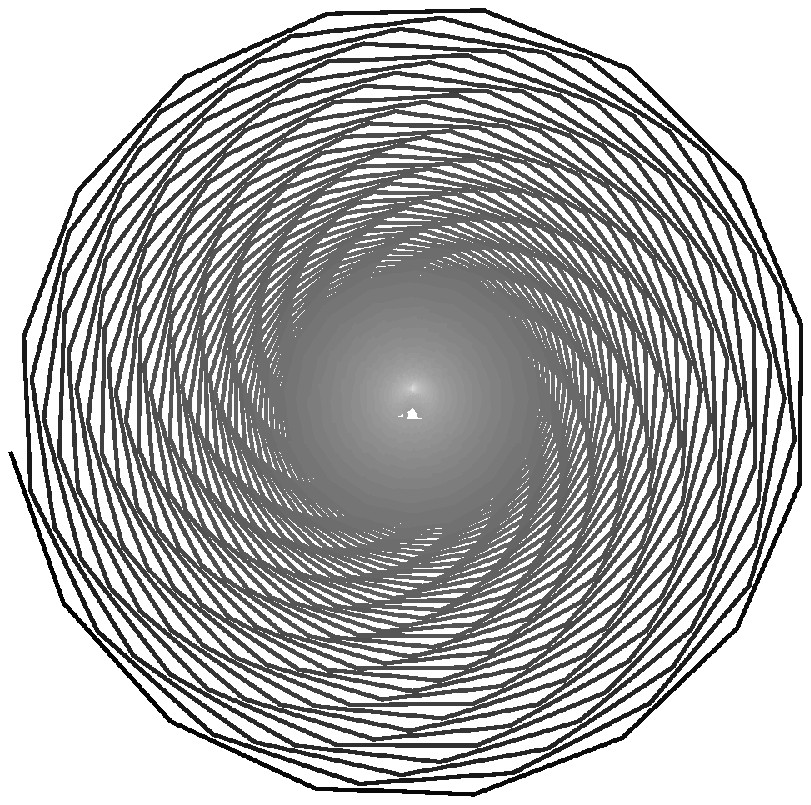}
}
\caption[]{Illustration of the (a) \emph{smiley}, (b) \emph{d-geom} and (c) \emph{ikeda} databases.}%
\label{fig:databases}
\end{figure}

We define three databases to study our identification algorithm. The \emph{smiley} test
has 2-dimensional source components \mbox{($d_m=2$)} generated from images of the 6 basic facial expressions\footnote{See http://www.smileyworld.com.}, see Fig.~\ref{fig:databases}(a). Sources $\bof{e}^m$ were generated by sampling 2-dimensional
coordinates proportional to the corresponding pixel intensities. In other words, the 2-dimensional images were
considered as density functions. $M\le 6$ was chosen. In the \emph{d-geom} dataset $\bof{e}^m$s were random variables uniformly
distributed on \mbox{$d_m$-dimensional} geometric forms. Geometrical forms were chosen as follows. We used: (i) the surface of the unit ball, (ii) the straight lines that connect the opposing corners of the unit cube, (iii) the broken line between $d_m+1$ points $\bof{0}\rightarrow\bof{e}_1\rightarrow\bof{e}_1+\bof{e}_2\rightarrow\ldots\rightarrow\bof{e}_1+\ldots+\bof{e}_{d_m}$ (where
$\bof{e}_i$ is the $i$ canonical basis vector in $\R^{d_m}$, i.e., all of its coordinates are zero except the $i$, which is 1), and (iv) the skeleton of the unit square. Thus, the number of components $M$ was equal to $4$, and the dimension of the components ($d_m$) can be different and scaled.  For illustration, see Fig.~\ref{fig:databases}(b). In the \emph{ikeda} test, hidden $\bof{s}^m_t=[s^m_{t,1},s^m_{t,2}]\in\R^2$ sources realized the ikeda map
\begin{align}
s^m_{t+1,1} &= 1 + \lambda_m[s^m_{t,1}\cos (w^m_t) - s^m_{t,2}\sin (w^m_t)],\\
s^m_{t+1,2} &= \lambda_m[s^m_{t,1}\sin (w^m_t) + s^m_{t,2}\cos (w^m_t)],
\end{align}
where $\lambda_m$ is a parameter of the dynamical system and $w^m_t=0.4-\frac{6}{1+(s^m_{t,1})^2+(s^m_{t,2})^2}$. $M=2$ was chosen with initial points $\bof{s}^1_1=[20;20]$, $\bof{s}^2_1=[-100,30]$ and parameters $\lambda_1=0.9994$, $\lambda_2=0.998$, see Fig.~\ref{fig:databases}(c) for illustration.

\subsection{Performance Measure, the Amari-index}\label{sec:amari-index}
The identification of the fAR-IPA model is ambiguous, the hidden $\bof{s}^m$ sources can be estimated up to ISA ambiguities. These ambiguities are however simple \cite{theis06towards}: the components of equal dimension can be recovered up to the permutation (of equal dimension) and invertible transformation within the subspaces. Thus, in the ideal case, the product of the ISA demixing matrix $\bof{W}_{\text{ISA}}$ and the ISA mixing matrix $\bof{A}$, $\bof{G}=\bof{W}_{\text{ISA}}\bof{A}$ is a block-permutation matrix. This property can be measured by a simple extension of the Amari-index \cite{amari96new}. Namely, one can (i) assume without loss of generality that the component dimensions and their estimations are ordered in increasing order ($d_1\le\ldots\le d_M$, $\hat{d}_1\le\ldots\le \hat{d}_M$),  (ii) decompose $\bof{G}$ into $d_i\times d_j$ blocks ($\bof{G}=\left[\bof{G}^{ij}\right]_{i,j=1,\ldots,M}$) and define $g^{ij}$ as the sum of the absolute values of the elements of the matrix $\bof{G}^{ij}\in\R^{d_i\times d_j}$. Then the Amari-index adapted to the ISA task of different component dimensions is defined as
\begin{equation}
    r(\bof{G}):=\frac{1}{2M(M-1)}\left[\sum_{i=1}^M\left(\frac{
\sum_{j=1}^Mg^{ij}}{\max_jg^{ij}}-1\right)+
\sum_{j=1}^M\left(\frac{
\sum_{i=1}^Mg^{ij}}{\max_ig^{ij}}-1\right)\right].\label{eq:Amari-index:def}
\end{equation}
One can see that $0\le r(\bof{G})\le 1$ for any matrix $\bof{G}$, and
$r(\bof{G})=0$ if and only if $\bof{G}$ is block-permutation matrix with $d_i\times d_j$ sized blocks. $r(\bof{G})=1$ is in the worst case, i.e, when all the $g^{ij}$ elements are equal in absolute value.

\subsection{Simulations}\label{sec:simulations}
Results on databases \emph{smiley}, \emph{d-geom} and \emph{ikeda} are provided here. For illustration purposes, we chose fAR order $p=1$ and used the recursive Nadaraya-Watson \eqref{eq:recNW} for functional AR estimation with the Gaussian kernel. The ISA subproblem was solved on the basis of the ISA separation theorem: the estimated ICA elements were clustered. The kernel canonical correlation technique \cite{bach02kernel} was applied to estimate the dependence of the ICA elements. The permutation search (clustering step) was carried out by greedy optimization for tasks of known component dimensions (\emph{smiley}, \emph{d-geom} datasets). We employed the NCut \cite{yu03multiclass} spectral technique on the \emph{ikeda} dataset to estimate unknown dimensions and to perform clustering. FastICA \cite{hyvarinen97fast} was used for the ICA estimation. Mixing matrix $\bof{A}$ was random orthogonal. For dataset \emph{smiley} and \emph{d-geom}, $\bof{f}$ was the composition of a random $\bof{F}$ matrix with entries distributed uniformly on interval $[0,1]$ and the noninvertible sine function. The Amari-index (Section~\ref{sec:amari-index}) was used to evaluate the performance of the proposed fAR-IPA method. For each individual parameter, $10$ random runs were averaged. Our parameters included $T$, the sample number  of observations $\bof{x}_t$, and bandwith $\beta \in (0,1/D)$ to study the robustness of the kernel regression approach. $\beta$ was reparameterized as $\beta = \frac{\beta_c}{D}$ and $\beta_c$ was chosen from the set $\{\frac{1}{2},\frac{1}{4},\frac{1}{8},\frac{1}{16},\frac{1}{32},\frac{1}{64}\}$. The performance of the method is summarized by notched boxed plots, which show the quartiles ($Q_1,Q_2,Q_3$), depict the outliers, i.e., those that fall outside of interval $[Q_1 - 1.5(Q_3-Q_1), Q_3 + 1.5(Q_3-Q_1)]$ by circles, and whiskers represent the largest and smallest non-outlier data points.

\begin{figure}[h]%
\centering%
\subfloat[][]{\includegraphics[width=6.3cm]{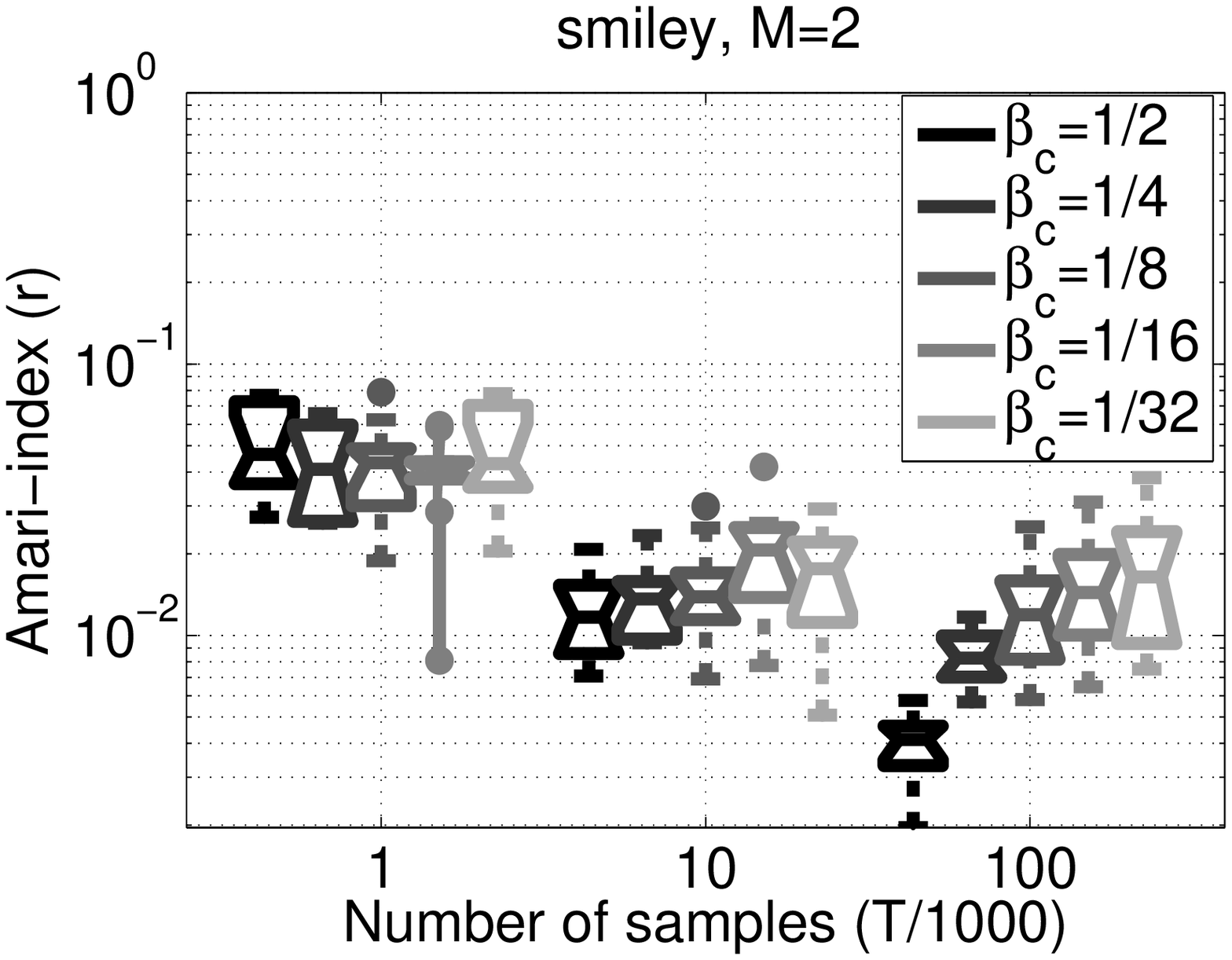}}%
\subfloat[][]{\includegraphics[width=6.3cm]{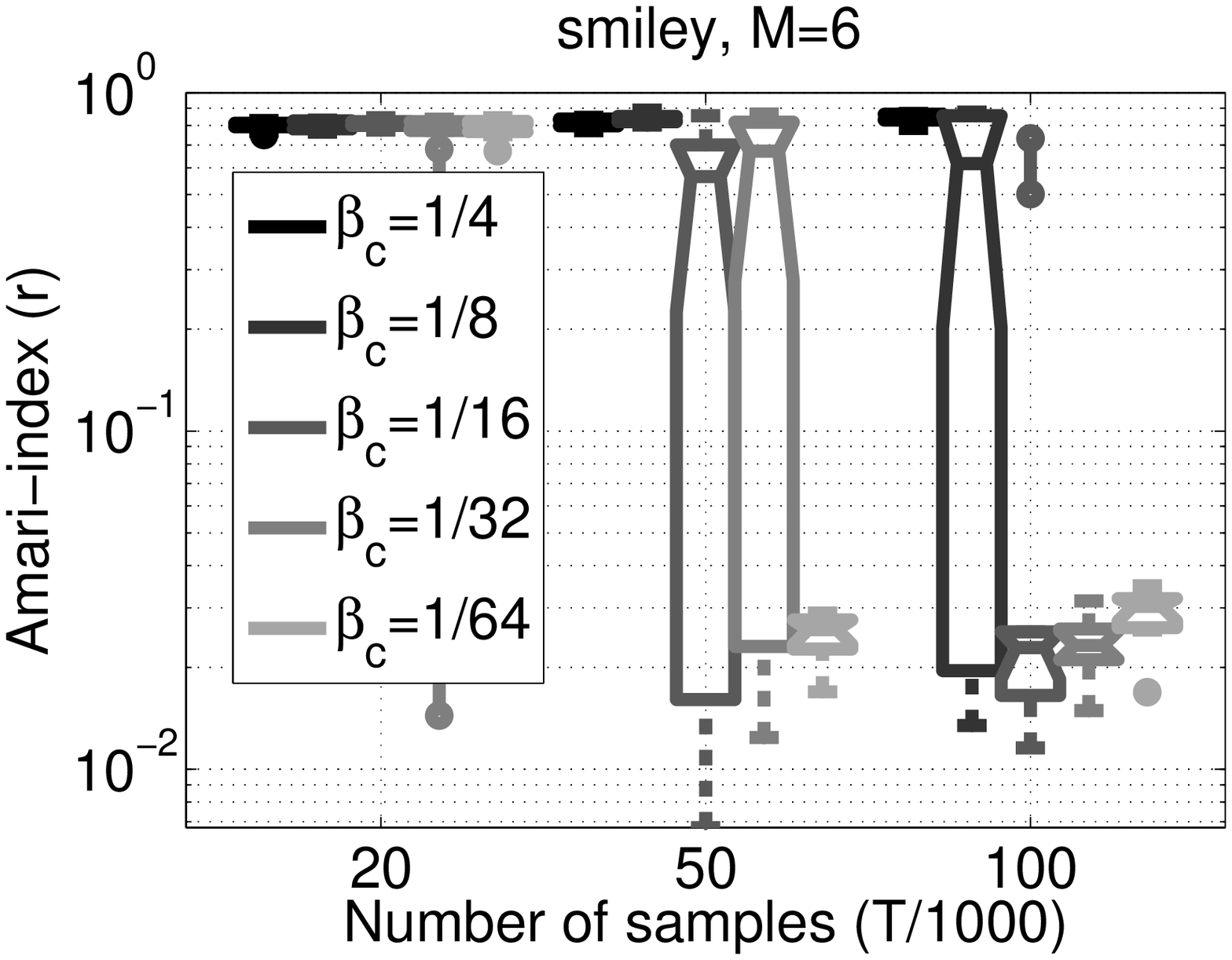}}\\%
\hfill\subfloat[][]{\includegraphics[width=4.4cm]{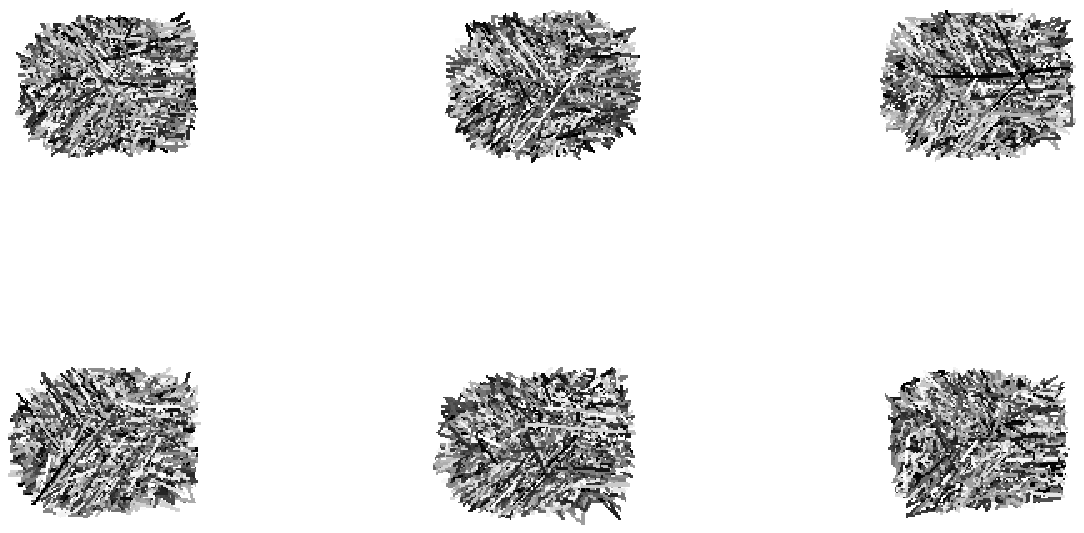}}\hfill
\subfloat[][]{\includegraphics[width=2cm]{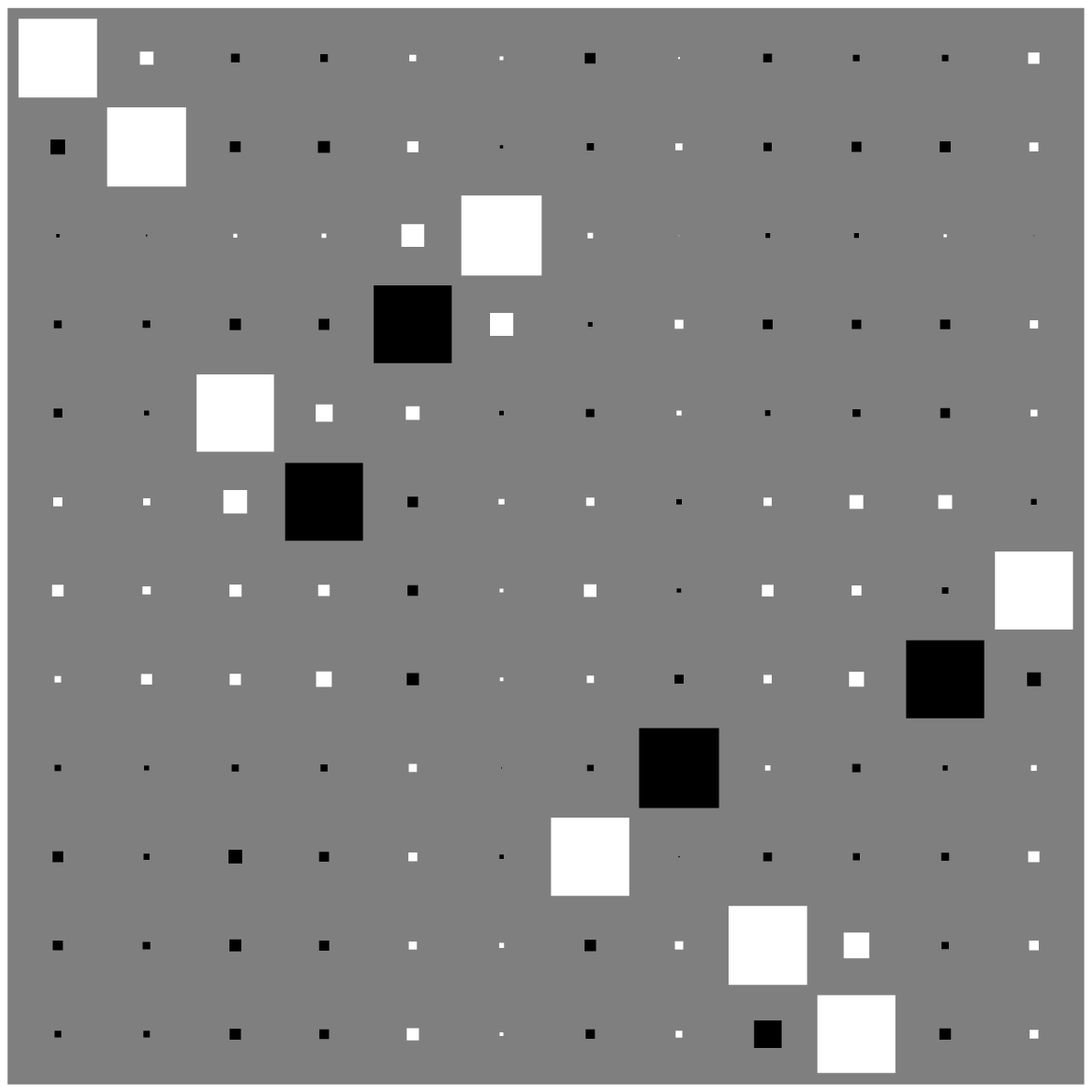}}\hfill\hfill
\subfloat[][]{\includegraphics[width=4cm]{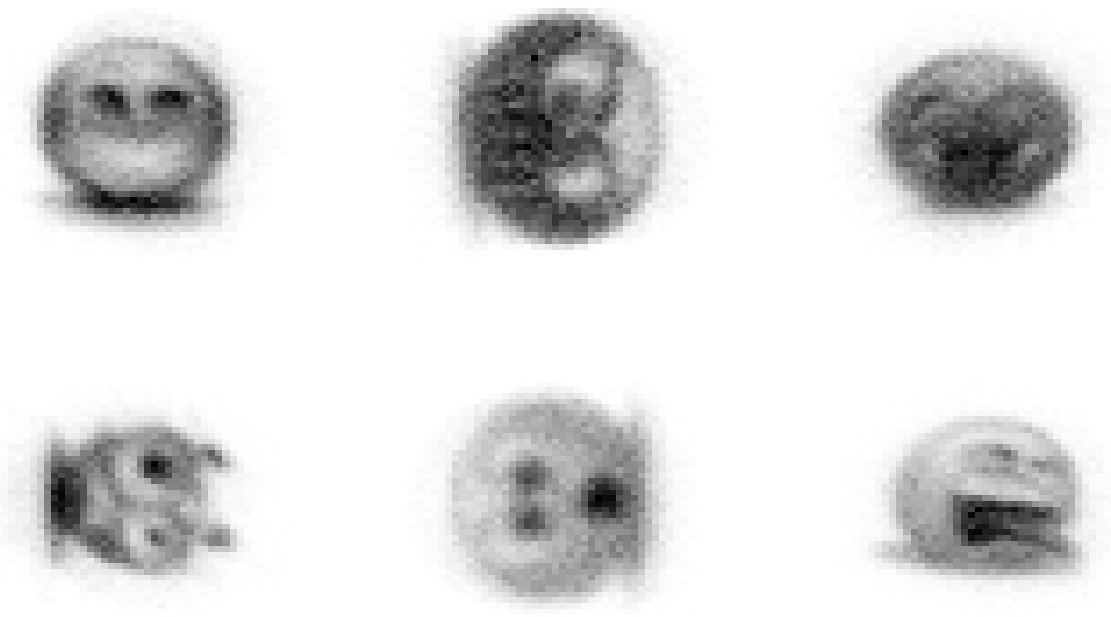}}\hfill\hfill
\caption[]{Illustration of the estimations on the \emph{smiley} dataset. (a)-(b): Amari-index as a function of the sample number, for $M=2$ and $M=6$ components, respectively. The estimation error is plotted on log scale for different bandwith parameters. (c): observed signal $\bof{x}$. (e): estimated components ($\hat{\bof{e}}^m$) with average (closest to the median) Amari-index for $M=6$, $\beta_c=\frac{1}{32}$, $T=100,000$. (d): Hinton-diagram of matrix $\bof{G}$ for (e)--it is approximately a block-permutation matrix with $2\times 2$ blocks.}%
\label{fig:demo:smiley}
\end{figure}

Figure~\ref{fig:demo:smiley} demonstrates that the algorithm was able to uncover the hidden components with high
precision for the \emph{smiley} dataset. Figure~\ref{fig:demo:smiley}(a) illustrates the $M=2$ ($D=4$) case, Fig.~\ref{fig:demo:smiley}(b) indicates that the problem with $M=6$ components ($D=12$) for $T=50,000-100,000$ samples is still amenable to the method. According to the figures, the estimation is robust with respect to the choice of bandwith.
The obtained source estimations are illustrated in  Fig.~\ref{fig:demo:smiley}(c)-(e).

Our experiences concerning the \emph{d-geom} and the \emph{ikeda} datasets are summarized in Fig.~\ref{fig:demo:d-geom,ikeda}. In accordance with the \emph{smiley} test, the dimension of the \emph{d-geom} problem was $D=12$, however the dimensions of the hidden components were different and unknown to the algorithm: $d_1=2$, $d_2=d_3=3$, $d_4=4$. As it can be seen from Fig.~\ref{fig:demo:d-geom,ikeda}(a) the method provides precise estimations on the \emph{d-geom} database for sample number $T=100,000-150,000$. Hinton-diagram of matrix $\bof{G}$ with average (closest to the median) Amari-index is depicted in Fig.~\ref{fig:demo:d-geom,ikeda}(c). Our third example is the \emph{ikeda} database. As it is illustrated in Fig.~\ref{fig:demo:d-geom,ikeda}(b), in this case an autoregressive approximation (AR-IPA) could not find the proper subspaces. Nevertheless, the Amari-index values of Fig.~\ref{fig:demo:d-geom,ikeda}(b) show that a functional AR-IPA approach was able to recover the hidden subspaces, for sample number $T\ge 10,000$. The figure also shows that the estimation is precise for a wide range of bandwith parameters. Hidden sources with average Amari-index uncovered by the method are illustrated Fig.~\ref{fig:demo:d-geom,ikeda}(d)-(f).

\begin{figure}[h]%
\centering%
\subfloat[][]{\includegraphics[width=6.3cm]{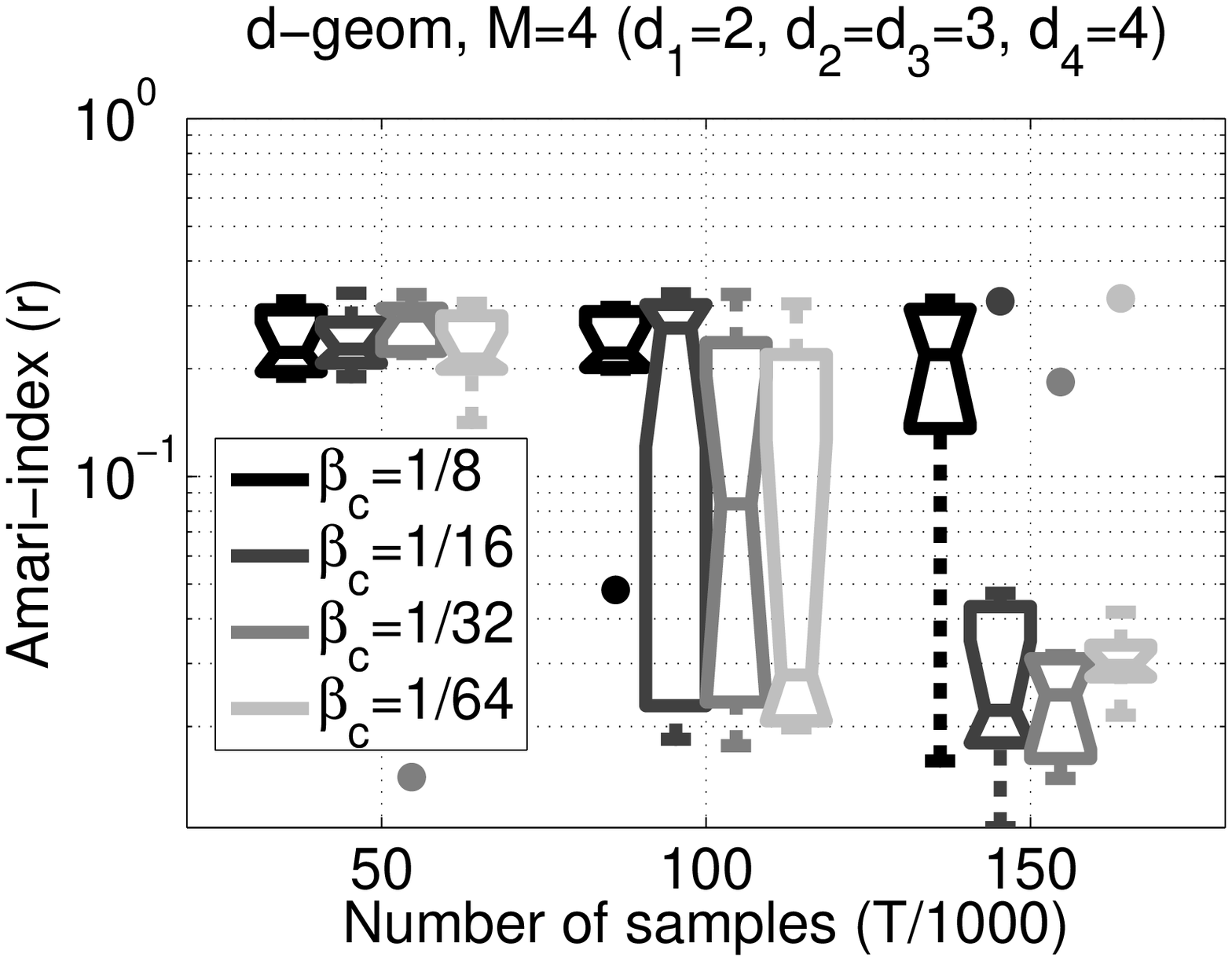}}%
\subfloat[][]{\includegraphics[width=6.3cm]{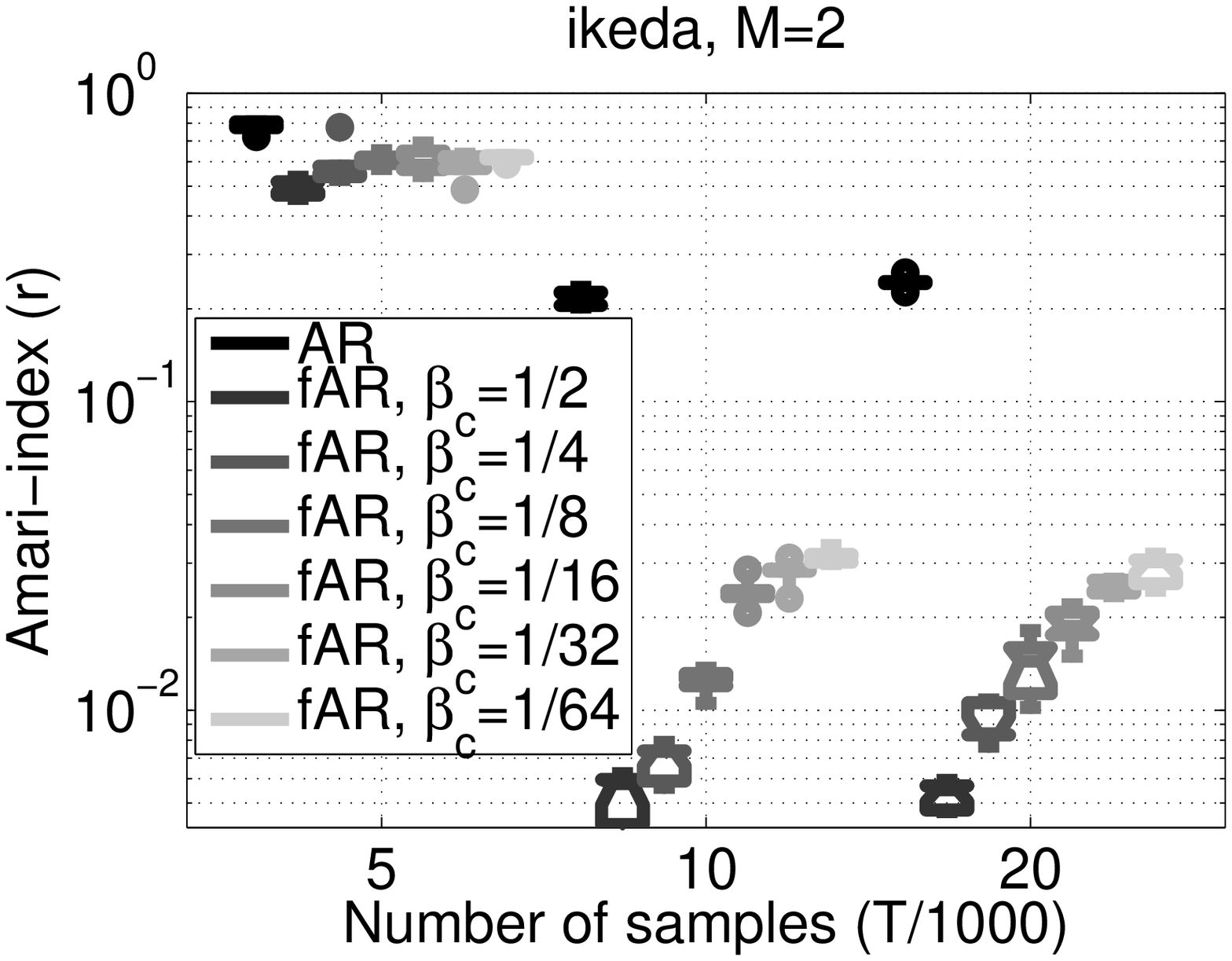}}\\
\subfloat[][]{\includegraphics[width=2.1cm]{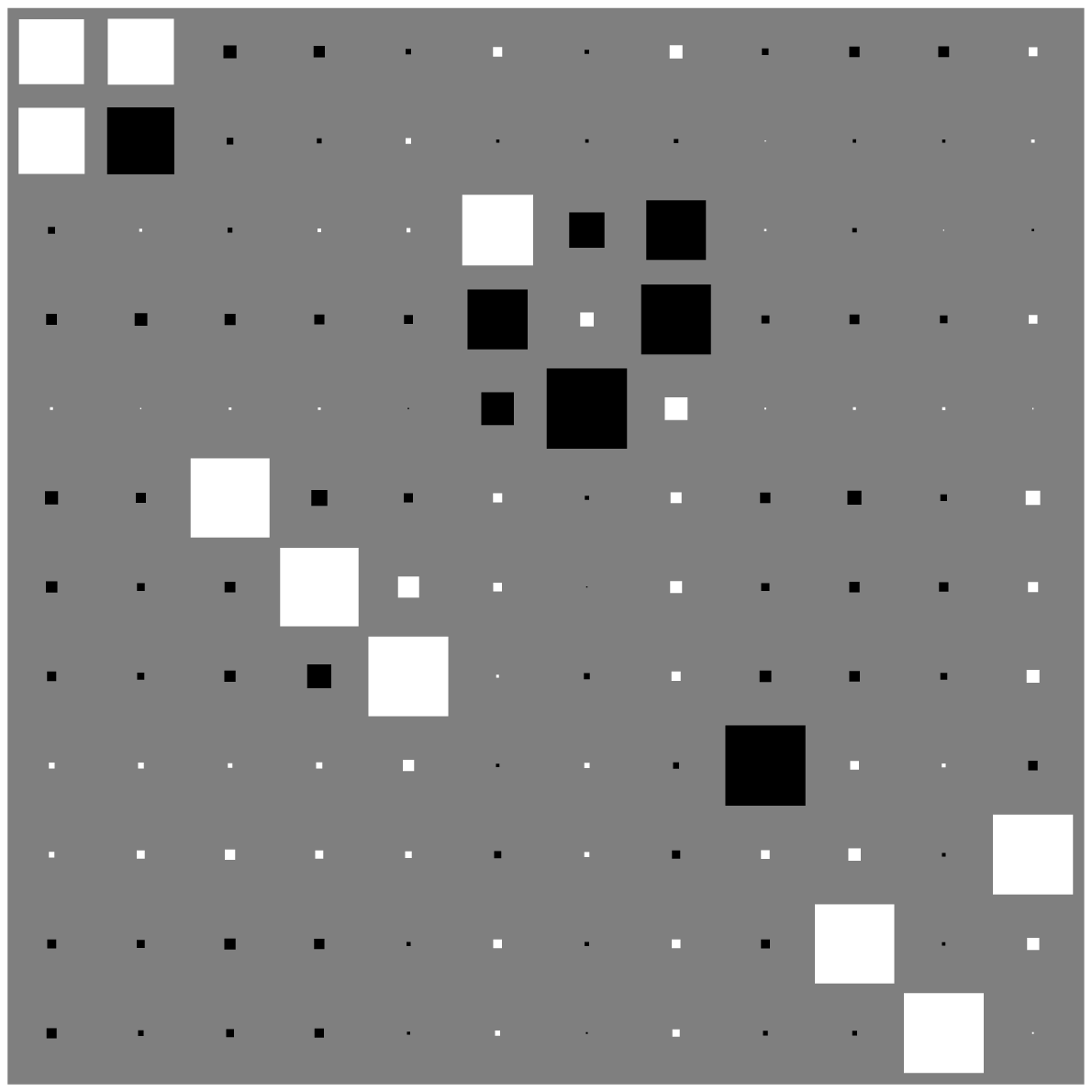}}\hspace*{0.05cm}
\subfloat[][]{\includegraphics[height=2.3cm]{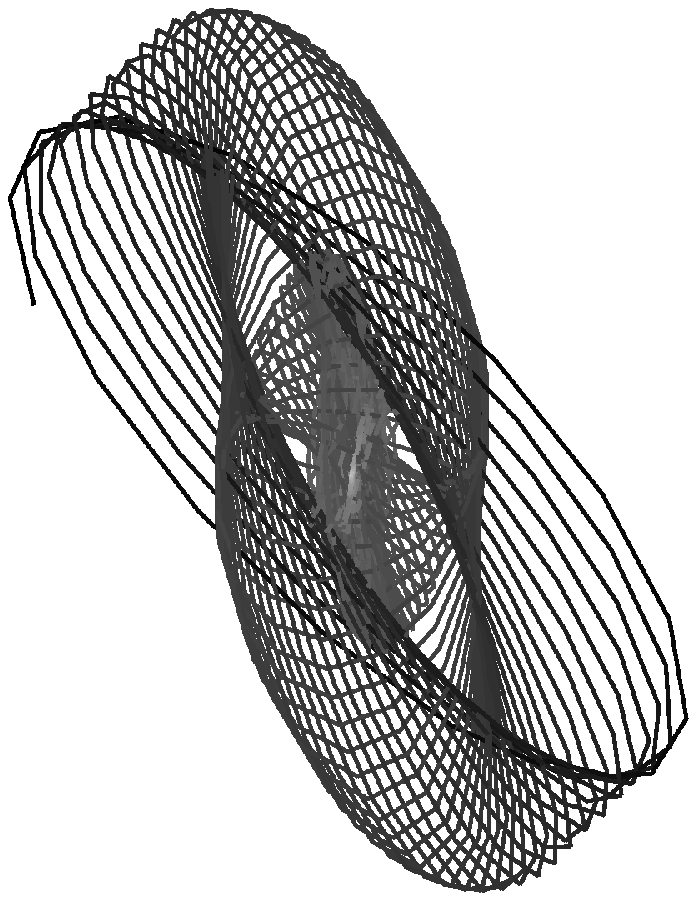}
\includegraphics[height=2.3cm]{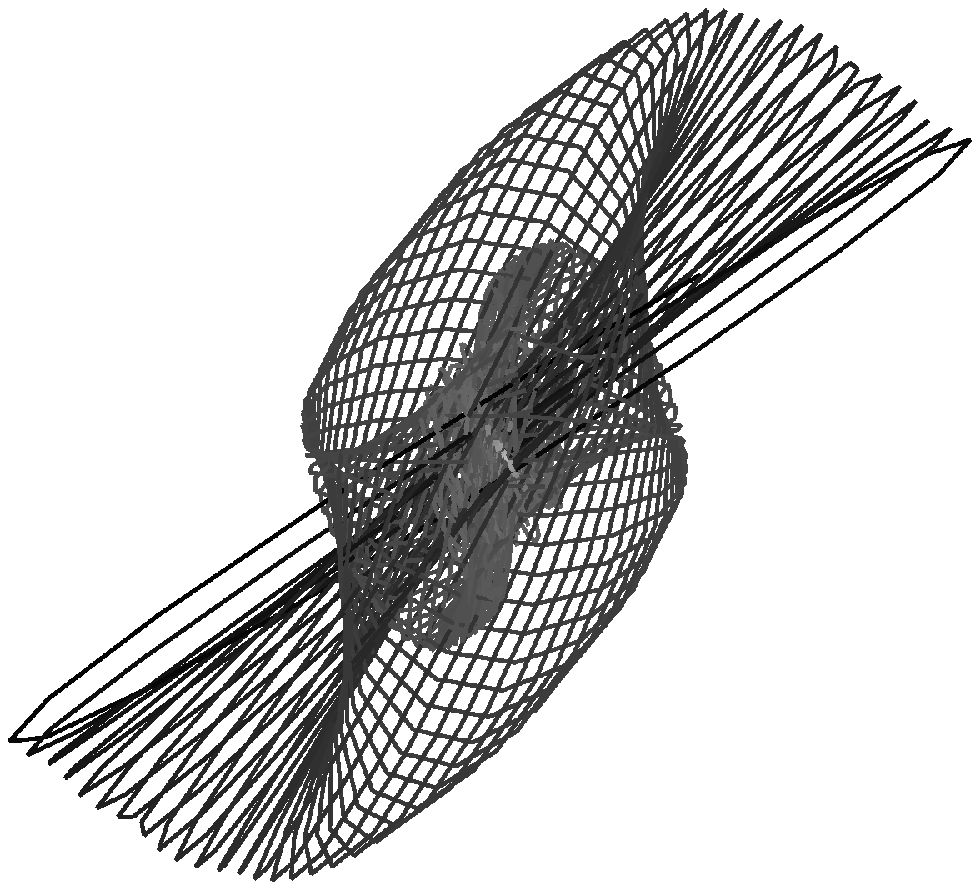}}\hspace*{0.05cm}
\subfloat[][]{\includegraphics[width=1.8cm]{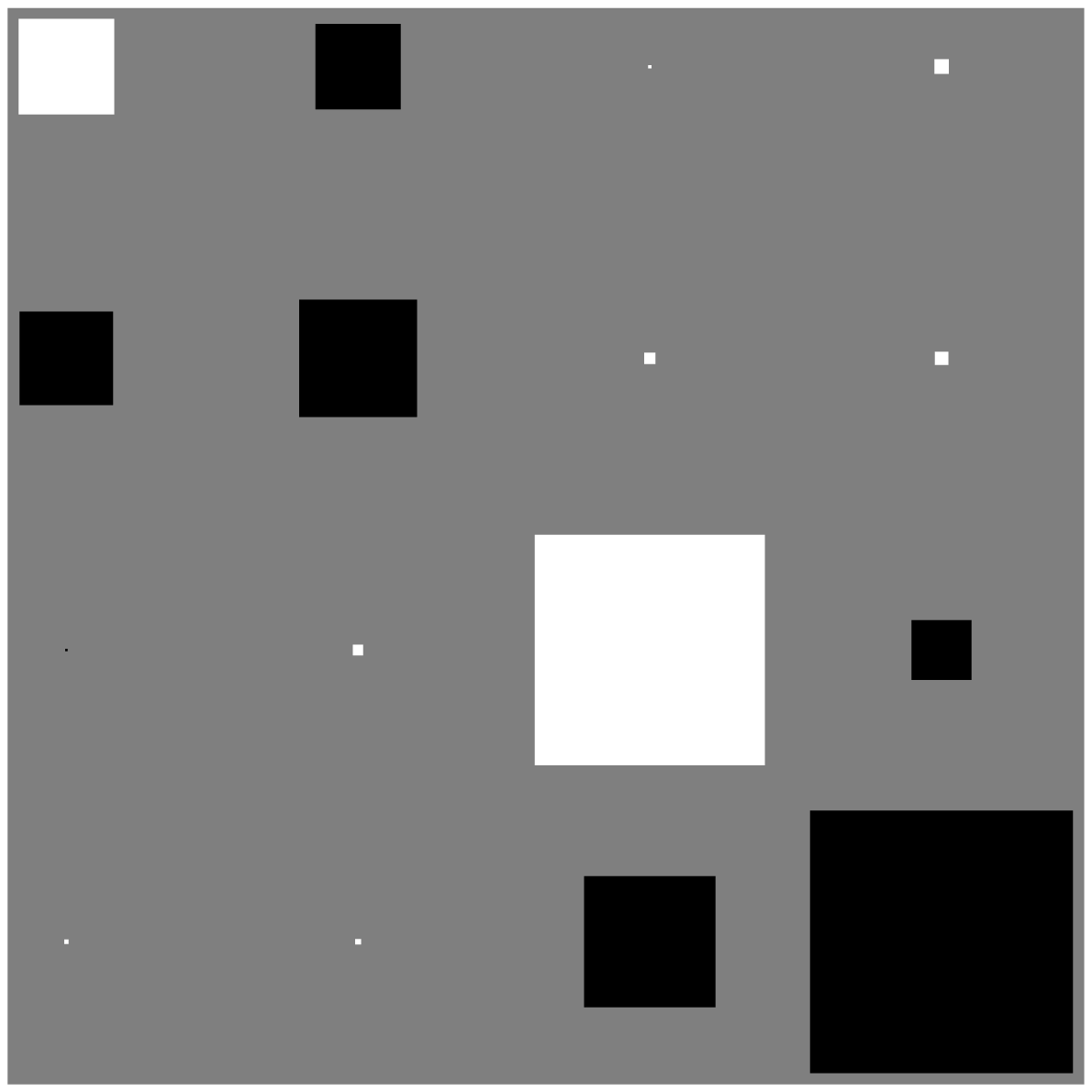}}\hspace*{0.05cm}
\subfloat[][]{\includegraphics[height=2.3cm]{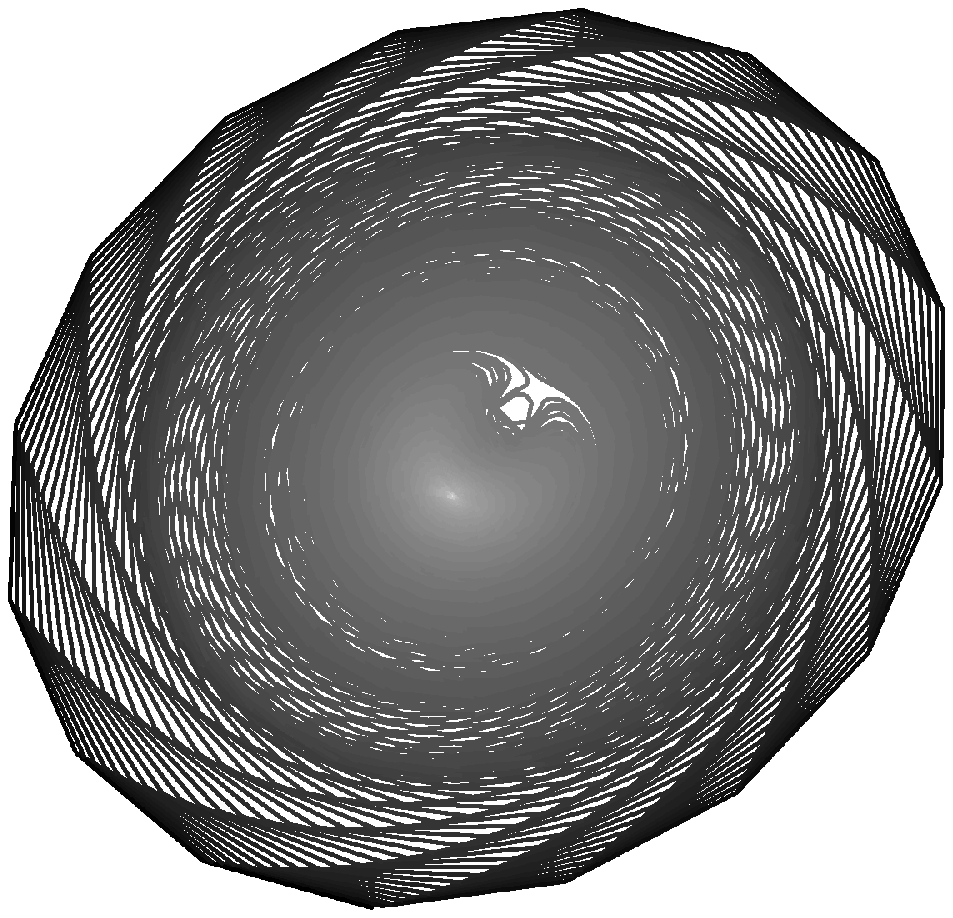}
\includegraphics[height=2.3cm]{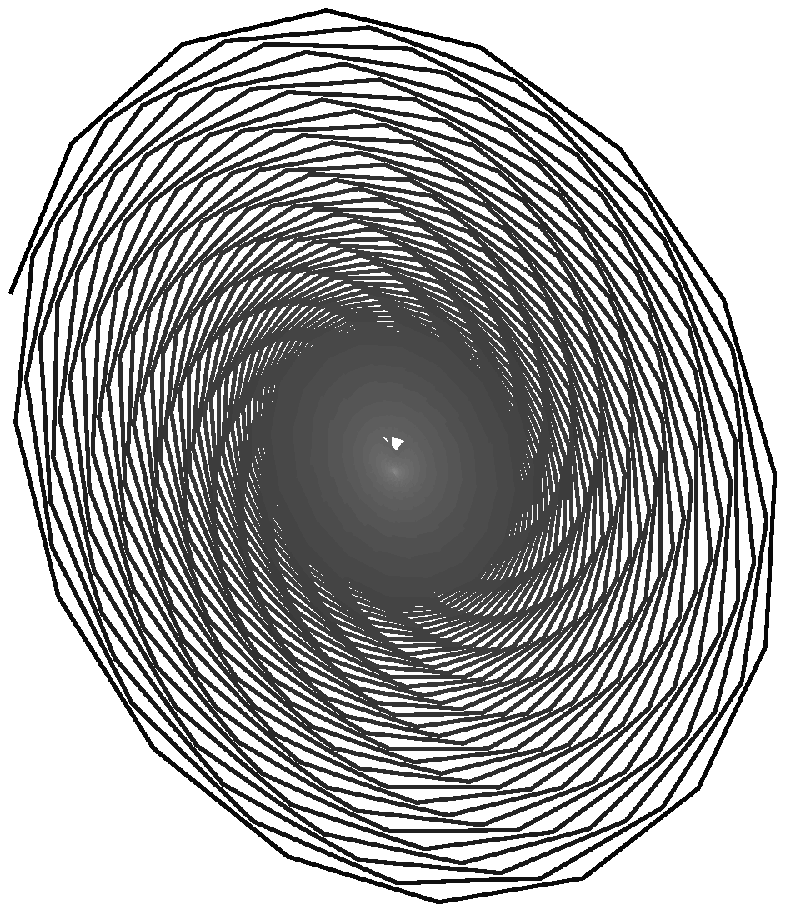}}
\caption[]{Illustration of the estimations on the \emph{d-geom} and the \emph{ikeda} dataset. (a)-(b): Amari-index on log scale as a function of the sample number for different bandwith parameters on the \emph{d-geom} (with component dimensions: $d_1=2$, $d_2=d_3=3$, $d_4=4$) and the \emph{ikeda} database, respectively. (c): Hinton-diagram of $\bof{G}$ with average (closest to the median) Amari-index for dataset \emph{d-geom}, $\beta_c=\frac{1}{32}$, $T=150,000$--it is approximately a block-permutation matrix with one $2\times 2$, two $3\times 3$ and one $4\times 4$ block. (d)-(f): estimation with average Amari-index for database \emph{ikeda}, $\beta_c=\frac{1}{2}$, $T=20,000$. (d): observation $\bof{x}$. (f): estimated components ($\hat{\bof{s}}^m$). (e): Hinton-diagram of matrix $\bof{G}$ for (f)--it is approximately a block-permutation matrix with $2\times 2$ blocks.}%
\label{fig:demo:d-geom,ikeda}
\end{figure}

\section{Conclusions}\label{sec:conclusions}
In this paper we (i) extended independent subspace analysis (ISA) to the not strictly stationary domain, (ii) relaxed the constraint of decoupled (block-decorrelated) dynamics, and (iii) simultaneously addressed the case of unknown source component dimensions. The temporal evolution of the sources was captured by functional autoregressive (fAR) processes. We generalized the ISA separation technique to the derived fAR setting (fAR-IPA, IPA-independent process analysis) and reduced the solution of the problem to fAR identification and ISA. The fAR estimation was carried out by the Nadaraya-Watson kernel regression method with strong consistency guarantee. We extended the Amari-index to different dimensional components and illustrated our technique by numerical experiments. According to the experiences, the fAR-IPA identification can be accomplished robustly and can be advantageous compared to a parametric approach. The robustness of the separation principle indicate that it can be extended
to a larger class of processes.

\bibliographystyle{splncs}
\bibliography{main}

\end{document}